\documentclass[lettersize,journal]{IEEEtran}

\usepackage{tabularx}

\ifCLASSINFOpdf
\else
\fi
\usepackage[numbers]{natbib}
\usepackage{epsfig}
\usepackage{epstopdf}
\usepackage[T1]{fontenc}
\usepackage{amssymb}
\usepackage{amsmath}
\usepackage{amsfonts}
\usepackage{verbatim}
\usepackage{graphicx}
\usepackage{hyperref}
\usepackage[usenames,dvipsnames]{xcolor}
\usepackage{lipsum}

\usepackage{multicol}
\usepackage{pgfplots}
\usepackage{lipsum}
\usepackage{colortbl}
\usepackage{caption}
\usepackage{algorithm}
\usepackage[noend]{algpseudocode}
\renewcommand{\algorithmicthen}{}
\usepackage{enumitem}
\makeatletter
\def\algbackskip{\hskip-\ALG@thistlm}
\makeatother

\usepackage{multirow}

\usepackage{colortbl}
\usepackage{tabularx}
\usepackage{lipsum}  
\usepackage{pifont}
\usepackage{makecell}
\usepackage{float}
\usepackage{amsthm}
\usepackage{pgf}
\usepackage{array}
\usepackage{ragged2e} 
\usepackage{threeparttable}

\usepackage{booktabs} 
\usepackage[margin=2cm]{geometry}

\begin{document}
%
\title{Trojan-Resilient NTT: Protecting Against Control Flow and Timing Faults on Reconfigurable Platforms}

\author{\IEEEauthorblockN{Rourab Paul$^1$, Krishnendu Guha$^{2}$, Amlan Chakrabarti$^{3}$}
\\
\IEEEauthorblockA{
Dept. of Computer Science and Engineering, Shiv Nadar University, Chennai, Tamil Nadu, India$^1$\\
School of Computer Science and Information Technology, University College Cork, Ireland$^{2}$\\
School of IT, University of Calcutta, West Bengal, India$^{3}$\\
rourabpaul@snuchennai.edu.in$^1$, kguha@ucc.ie$^2$, acakcs@caluniv.ac.in$^3$} 
}

\maketitle
\vspace{-10pt}

\begin{abstract}
Number Theoretic Transform (NTT) is the most essential component for polynomial multiplications used in lattice-based Post-Quantum Cryptography (PQC) algorithms such as Kyber, Dilithium, NTRU etc. However, side-channel attacks (SCA) and hardware vulnerabilities in the form of hardware Trojans may alter control signals to disrupt circuit’s control flow and introduce unconventional delays in the critical hardware of PQC. Hardware Trojans, especially on control signals, are more low cost and impactful than data signals because a single corrupted control signal can disrupt or bypass entire computation sequences, whereas data faults usually cause only localized errors. On the other hand, adversaries can perform Soft Analytical Side Channel Attacks (SASCA) on the design using the inserted hardware Trojan. In this paper, we present a secure NTT architecture capable of detecting unconventional delays, control-flow disruptions, and SASCA, while providing an adaptive fault-correction methodology for their mitigation. 
Extensive simulations and implementations of our {\sf Secure NTT} on Artix-7 FPGA with different Kyber variants show that our fault detection and correction modules can efficiently detect and correct faults whether caused unintentionally or intentionally by hardware Trojans—with a high success rate, while introducing only modest area and time overheads.
\end{abstract}
\vspace{-10pt}
\section{Introduction}
The NTT significantly reduces the time complexity of multiplying degree-\( n-1 \) polynomials from \( \mathcal{O}(n^2) \) to \( \mathcal{O}(n \log n) \). As a result, standardized lattice-based PQC schemes such as Kyber, Dilithium, and NTRU benefit from substantial acceleration across various secure processing platforms. This acceleration of PQCs \cite{li}, \cite{bin}, \cite{mert} in FPGA makes them suitable for future high-speed data communication infrastructure. However, NTT suffers from side-channel attacks \cite{primas} and other hardware Trojans \cite{jati} that may reveal multiple secret shares during NTT computation. A hardware Trojan \cite{Ghandali} can deliberately induce controlled side-channel leakages like SASCA \cite{ravi2}, \cite{rafael}. The growing demand for FPGAs in defense, aerospace, automotive, and telecommunications is projected to reach 11 billion USD \cite{fpga_market} by 2027. This rising popularity has globalized the FPGA supply chain, thereby increasing the risk of hardware Trojans in modern systems. Consequently, the attack surface for hardware Trojan insertion has expanded across multiple stages, including RTL \cite{dai}, synthesis \cite{mukhrejee}, or foundry stages. Moreover, hardware Trojans can also be inserted after bitstream generation \cite{bit_alter:pawel}. The variety of Trojan insertion stages and the diverse nature of hardware Trojans mainly target the data signals and control signals of the system. However, attackers mostly \cite{wu} prefer inserting Trojans into control signals due to their low cost and high impact. For example, gating the clock, flipping the reset, or delaying the enable signal can stall or misdirect the entire datapath and disrupt the control flow.
\vspace{-10pt}
\subsection{Threat Models}
In FPGA fabric, vulnerabilities can arise intentionally or unintentionally. Unintentional vulnerabilities may result from aging that may cause performance degradation and excessive power consumption over time. Additionally, adversaries present in third-party design houses may intentionally insert hardware Trojans into the FPGA fabric (design-time or foundry-level) or bitstream (post-design, deployment-time). Hardware Trojans may be introduced into the system through compromised CAD tools \cite{zhang}. A malicious CAD tool could bypass or neutralize testing and post-silicon validation \cite{basu}. Trojans can also be inserted during deployment by tampering with the bitstream \cite{bit_alter:pawel}. On the other hand, without any modification of the design, purely observational attacks such as SASCA \cite{ravi2}, \cite{rafael} can be extremely powerful against cryptographic systems if is unprotected. All these attack phases may affect the integrity, confidentiality and availability of the crypto system. In this work, we focus on hardware vulnerabilities that cause unconventional delays, disrupt the expected behavior (Control Flow) of the hardware, cause information leakage, produce erroneous outputs and degrade overall circuit performance.  
If the proposed {\sf Secure NTT} is part of a malicious bitstream, or placed in an infected FPGA fabric, or subjected to aged or heated FPGA conditions, any unconventional delays or abnormal control flow in the NTT caused by these factors can be detected.
\subsection{Literature}
The existing literature on NTT can broadly be categorized into two directions. The first line of work focuses primarily on optimizing NTT implementations for ASICs, FPGAs, and embedded processors \cite{bin}, \cite{li}, \cite{mert}. In contrast, another body of research addresses the protection of NTT against hardware Trojans and side-channel attacks such as SASCA. Prior works on protected NTT have mainly proposed fault-detection mechanisms, targeting either data signals or control signals.
\subsubsection{Protection on Data Signals of NTT}
To protect the data signals of the NTT, it is necessary to secure the arithmetic operations of the NTT, since these operations are the source of the data. As shown in Algorithm \ref{algo:intt}, the main data signals of the NTT involve the computation of $V$ through polynomial multiplication followed by modular reduction, as well as the memory addresses $j$ and $k$ used to access the coefficient memory and twiddle-factor memory, respectively. Article \cite{sarker} proposed a fault detection module for NTT which targets the polynomial multiplication process of the NTT. They recomputed the multiplication result using the Recomputation with Negate Operands (RENO) technique on Spartan-6 and Zynq UltraScale+ platforms, incurring a 12.74\% resource overhead and approximately 20\% power overhead. The proposed RENO technique can detect errors in $A[k1]$ and $\omega$ (Line \ref{line:mrfd} of Algorithm \ref{algo:intt}) during the polynomial multiplication and reduction operations. Similarly, \cite{paul} presents a fault-detection model for the same $V$ computation using Recomputation with Modular Offset (REMO), combined with a word-wise Montgomery reduction, implemented on an Artix-7 FPGA with an 8.5\% slice overhead and a 1.8\% energy overhead. They identified a correlation among the indices $i$, $j$, and $k$ (Algorithm \ref{algo:intt}), which generate the memory addresses for polynomial coefficients and twiddle factors memories in each NTT iteration (Algorithm \ref{algo:intt}). This correlation-based method can detect hardware Trojans inserted in the memory address generation logic.
 Article \cite{agha} has also targeted Montgomery Reduction of the NTT using Recomputation based Shifted Operand (RESO). Sven et al. \cite{sven} target the data signals of the NTT, employing interpolation, evaluation, and inverse NTT techniques to detect faults in the multiplication with twiddle factors and the addition operations on the ARM Cortex-M4 platform. 
\subsubsection{Protection on Control Signals of NTT} Except for the aforementioned fault occurrences on data signals of the NTT, some works in the literature address control-flow and delay attacks caused by disruptions of the NTT control signals.
 Jati et al. \cite{jati} implemented a configurable CRYSTALS-Kyber architecture that addresses circuit delay and control flow integrity ($CFI$) violations caused by hardware Trojans inserted in control signals. This design includes the NTT, employs a duplicate inverted logic state machine to verify the Finite State Machine (FSM) of the Kyber processor and utilizes a Clock Cycle Counter ($CCC$) in the instruction module to detect unconventional delays in the entire crypto processor. 
 Since SCA can also be induced by hardware Trojans, several recent works have exploited power and timing leakages in NTT for lattice-based PQC, particularly SASCA attacks, using probabilistic models \cite{primas}, \cite{ravi}. To prevent SCAs on the NTT, Jati et al. \cite{jati} randomized memory addresses for each butterfly layer instead of using a linear increment in the memory address. This is possible because the order of NTT operation does not matter in a single butterfly layer. Article \cite{ravi2} also implemented the same memory address randomization technique for NTT to prevent side-channel attacks on the ARM Cortex-M4 platform. Additionally, it applied input masking using twiddle factors to protect NTT against SASCA. Rafael et al. \cite{rafael} adopted the same input masking technique using twiddle factors in the NTT on a reconfigurable platform. However, instead of masking all inputs, they selectively masked only a subset and varied the number of masked inputs to observe the effect on power consumption.

To the best of our knowledge, our {\sf Secure NTT} is the first design to detect hardware Trojan–based control-signal attacks, prevent unconventional delays and control-flow disruptions, and mitigate SASCA through local masking, while applying fault-specific adaptive correction measures. It should be noted that our {\sf Secure NTT} does not perform fault detection or correction on data signals, unlike the approaches in \cite{sven}, \cite{sarker}, \cite{agha} and \cite{paul}.  The contributions of our article are summarized as :
\begin{itemize}
    \item Unlike conventional duplicate Control Status Registers (CSR), we introduce a lightweight shift-register–based backup CSR, fully independent of the NTT logic, to verify and ensure Control Flow Integrity ($CFI$) in the NTT.
    Our adopted Clock Cycle Counter ($CCC$) mechanism safeguards all critical control and status signals against unintended delays, The input masking of polynomial coefficients with the twiddle factors in our NTT can prevent SASCA.
    \item The proposed adaptive fault-correction mechanism for FPGA-based designs utilizes multiple backups of partial (PR) bitstreams corresponding to different FPGA floor regions. Our proposed fault correction algorithm evaluates a risk factor ($R_i$) for each PR bitstreams corresponding to the FPGA floors of all critical components avaialble in a system. Based on this $R_i$ and the nature of the detected faults, it applies appropriate countermeasures. We implemented this adaptive fault correction methods for our NTT to dynamically respond to the specific nature of each detected fault, ensuring resilient operation against diverse hardware Trojan effects 
    \item The proposed {\sf Secure NTT} is implemented on an Artix-7 FPGA with $5$ pipeline stages without incurring any timing overhead in fault detection. The implementation overhead of our fault detection and correction modules is minimal and competitive with existing NTT solutions.
\end{itemize}
The organization of this paper is described as follows. Section \ref{sec:fd} discusses our fault detection methods. The working principle of our fault correction methods is discussed in Section \ref{sec:fc}. The Results of the
implementation, conclusions and future scopes are presented in Section \ref{sec:rai}, Section \ref{sec:con} and \ref{sec:fs} respectively.
\section{Fault Detection Modules}
\label{sec:fd}
NTT primarily consists of 3 operations: memory read, arithmetic operations (mult, add, sub, modulus) and memory write. In our fault detection model, arithmetic operations in the NTT are masked using Local Masking ($LM$) at each clock cycle, while unintentional delays and control-flow integrity issues are addressed using the Clock Cycle Counter ($CCC$) and Control Flow Integrity ($CFI$), respectively.
\begin{algorithm}
  \caption{NTT Algorithm}
  \label{algo:intt}
  \begin{algorithmic}[1]
    \State \textbf{Input:} $A(x)$, $\omega$, $q$
    \State \textbf{Output:} $\overline{A}(x)$
    \State $hl=\frac{n}{2}$
    \For{$i = 0$ to $log_2(n)-1$}
      \For{$j = 0$ to $2^i - 1$}
        \For{$k = 0$ to $hl-1$ }
          \State $\omega_j = \omega[j]$
          \State $k0=k$
          \State $k1=k+hl$
          \State $U = A[k0]$  \label{line:u}
          \State $V = (A[k1] \times \omega_j) \% q$ \label{line:mrfd}
          \State $\overline{A}[k0] = (U + V) \% q$ \label{uplusv}
          \State $\overline{A}[k1] = (U - V) \% q$ \label{uminusv}
        \EndFor
      \EndFor
      \State $hl=\frac{hl}{2}$
    \EndFor
    \State \Return $\overline{A}$
  \end{algorithmic}
\end{algorithm}

\subsection{Control Flow Integrity}
The proposed NTT includes 8 subcomponents (Table \ref{tab:csr}); among them, only the $Bit$ $Reverser$ and $w\_mem$ operate without control signals. The remaining 6 use control signals for activation/reset and status signals like $done$ or $ready$ to indicate output readiness. This structured signaling framework ensures seamless coordination and efficient data flow across the NTT architecture.
The control Flow Integrity of proposed {\sf Secure NTT} is maintained with 3 strategies.
 \begin{table}[!htbp]
 	\centering
	\begin{tabular}{
	|>{\centering\arraybackslash}p{1.15cm}
	|>{\centering\arraybackslash}p{1.1cm}
	|p{5.1cm}|}
	\hline
	\textbf{Sub components}& \textbf{CSR Signals} & \textbf{Remarks} \\ \hline
        $poly\_mem$ & \texttt{rd\_en}, \texttt{wr\_en}, \texttt{polymem} \texttt{\_ce} & poly\_mem stores polynomial coefficients, rd\_en and wr\_en are used to read write from/to poly\_mem, poly\_mem\_ce is used to enable poly\_mem\\\hline
        $CTRL$   & \texttt{CTRL\_rst} & $CTRL$ generate all the required control signals of sub-components, \texttt{CTRL\_rst} is used to initiate the address values for $poly\_mem$ and $w\_mem$ \\\hline
        $U$ $Buffer$ & \texttt{uBuff} \texttt{\_rst} & $uv\_add$ and $uv\_sub$ needs $u$ and $v$ as inputs, $u$ comes directly from $poly\_mem$ but $v$ needs $2$ clock cycles to generate from barret reduction. Therefore, $u$ needs to buffer for $2$ clock cycles. \texttt{uBuff\_rst} is used to reset pipeline buffers which hold the value $u$ until $v$ computes in Barrett Reduction \\\hline
        $Barrett$ $Reduction$& \texttt{barrett} \texttt{\_rst}, \texttt{barrett} \texttt{\_strt}, \texttt{barrett} \texttt{\_done} & It compute $v$ $\times$ $\omega$ mod $q$. \texttt{barrett\_rst} is used initialized intermediate registers of barrett reduction,  \texttt{barrett\_strt} is used to start the barret reduction when date from $poly\_mem$ and $w\_mem$ are ready, \texttt{barrett\_done} starts $uv\_adder$ and $uv\_sub$. \\\hline
        $uv\_adder$ \& $uv\_sub$  & \texttt{uv\_rst}, \texttt{uv\_strt} & This blocks are required to add and subtract $u$ and $v$. \texttt{uv\_rst} make $uv\_adde$r \& $uv\_sub$ in reset condition, \texttt{uv\_strt} starts $uv\_adder$ \& $uv\_sub$ \\\hline
        $Bit\_$ $Reverser$ &  - & This combinational logic block is required to generate a reading address of $\omega$ from $w\_mem$. \\\hline
        $w\_mem$ &  - & This is a ROM to store $\omega$ and it is a combinational logic block \\\hline
\end{tabular}
 		\caption{Components \& CSR Signals of Our Secure NTT}
 	\label{tab:csr}	
 \end{table}
 
 \subsubsection{Control Status Register (CSR)}
\begin{figure}[!htb]
\centering
\includegraphics[width=0.45\textwidth]{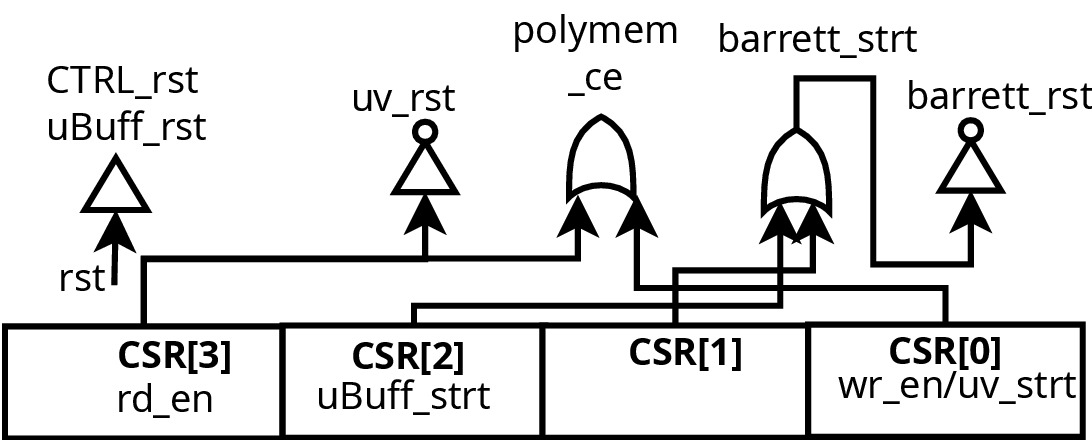}
\caption{Control and Status Signals of NTT }
\label{fig:NTT_csr}
\end{figure}

\begin{figure*}[!htb]
\centering
\includegraphics[width=0.7\textwidth]{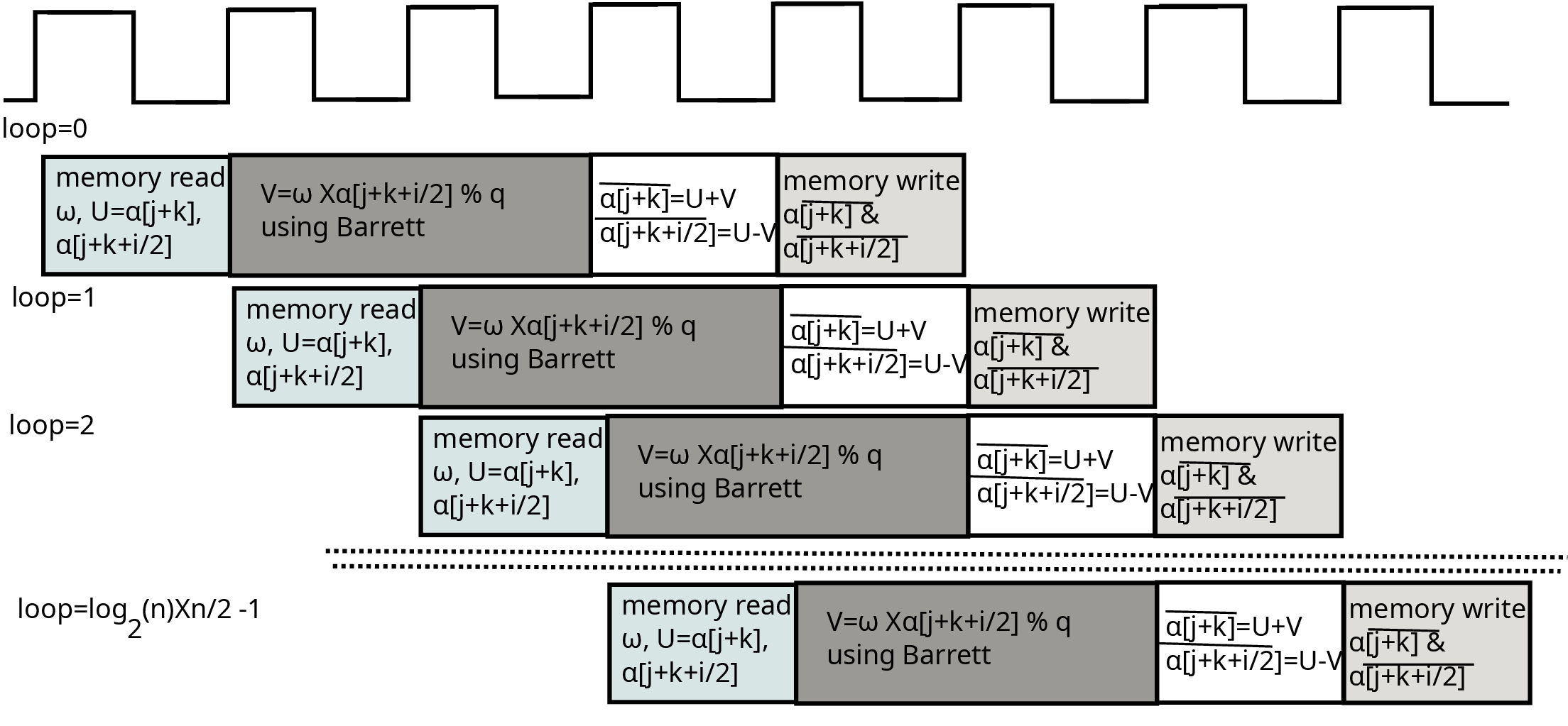}
\vspace{-5pt}
\caption{Pipeline Stages of proposed Secure NTT}
\vspace{-10pt}
\label{fig:NTT_pipe}
\end{figure*}

All input control signals for the sub-components of the proposed {\sf Secure NTT} are generated by a 4-bit Control Status Register ($CSR$) placed inside the $CTRL$ unit. The bits in $CSR$ change if the $rst$ input of the $NTT$ becomes low. The bits of $CSR$ from $\chi^{th}$ clock cycle to $(\chi{+}1)^{\text{th}}$ clock cycle can be represented as: 

{\scriptsize
\[
CSR_{i-1}(\chi{+}1) = 
\begin{cases}
CSR_{i}(\chi), & \text{if } 0 \leq i < \eta{-}2, \\
CSR_{i}(\chi), & \text{if } i = \eta{-}1,
\end{cases}
\text{for } \chi = 1,.., logn \times \frac{n}{2}
\]
}
Here $\chi$ indicates the clock cycles which iterates from $1$ to $logn \times \frac{n}{2}$ for the $NTT$ operation of $n-1$ degree polynomial. Therefore, the proposed $NTT$ requires $logn \times \frac{n}{2}$ clock cycles. Here, the $i^{\text{th}}$ bit of $CSR$ at the $\chi^{\text{th}}$ clock cycle, denoted as $CSR_i[\chi]$, shifts to the $(i{-}1)^{\text{th}}$ bit of $CSR$, i.e., $CSR_{i-1}[\chi+1]$, at the $(\chi{+}1)^{\text{th}}$ clock cycle. Therefore, this $CSR$ behaves like a right shift register. The size of the $CSR$ is $4$ bits because the proposed {\sf Secure NTT} has five pipeline stages as shown in Fig. \ref{fig:NTT_pipe}. Once the $rst$ signal is de-asserted, the first pipeline stage is activated, and the remaining four pipeline stages are controlled by the 4 bits of the $CSR$. In our case, $n=256$ and the pipeline depth is $p = 5$. Therefore, the complete $NTT$ operation requires $(\log n \times \frac{n}{2})+(p-1) =1024+4=1028$ clock cycles. The algorithm \ref{algo:intt} illustrates the NTT operation, where $A(x)$ is the input polynomial in coefficient form with $n = 256$ coefficients. The output $\overline{A(x)}$ represents the point-value form of $A(x)$, and $\omega$ denotes the twiddle factor.
 As shown in Fig. \ref{fig:NTT_csr}, the \texttt{CSR[3]} and \texttt{CSR[2]} bits are used to generate \texttt{rd\_en} and \texttt{uBuff\_rst} respectively. The \texttt{CSR[0]} is used for \texttt{wr\_en} and \texttt{uv\_strt}. The other required $CSR$ signals can be generated from $CSR$ using below equations.
\[
\hspace{-0.5em}
\begin{array}{l}
\small{\texttt{uv\_rst}=\lnot \texttt{CSR[3]}, 
\texttt{barrett\_strt}=\texttt{CSR[1]}\lor\texttt{CSR[2]},} \\
\small{\texttt{barrett\_rst} = \lnot (\texttt{CSR[1]} \lor \texttt{CSR[2]}),} \\
\small{\texttt{poly\_me\_ce} = \texttt{CSR[0]} \lor \texttt{CSR[3]}}

\end{array}
\]

\begin{figure*}[!htb]
\centering
\vspace{-5pt}
\includegraphics[width=0.75\textwidth]{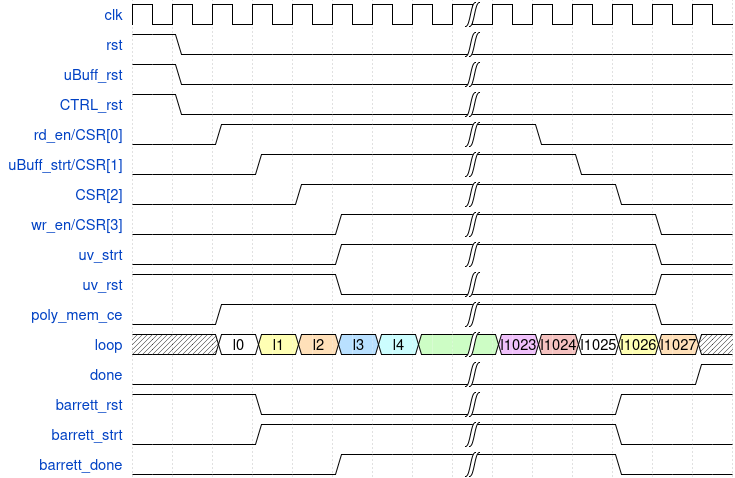}
\vspace{-5pt}
\caption{Timing Diagram of Our Secure NTT}
\label{fig:NTT_time}
\vspace{-10pt}
\end{figure*}

The reset inputs of $CTRL$ and $uBuff$ are \texttt{CTRL\_rst} and \texttt{uBuff\_rst} respectively which are directly generated from $rst$ input of the $NTT$. Therefore, \texttt{CTRL\_rst=rst} and \texttt{uBuff\_rst=rst}.
The $CSR$ is the core of the entire $NTT$, controlling all sub components by generating the necessary control signals. The timing diagram of the proposed $NTT$, along with all control and status signals, is shown in Fig. \ref{fig:NTT_time}.

  \begin{figure}[!htbp]
\centering
\vspace{-10pt}
\includegraphics[width=0.45\textwidth]{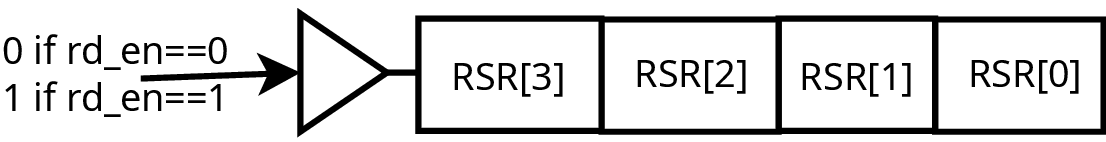}

\caption{Right Shift Register}
\label{fig:shift_reg}
\end{figure}

 \subsubsection{Right Shift Register (RSR)}
As shown in Fig. \ref{fig:NTT_time}, it is observed that after $rst$ of $NTT$ goes low, \texttt{rd\_en} (\texttt{CSR[3]}) becomes high. Subsequently, in the next three clock cycles, \texttt{uBuff\_rst} (\texttt{CSR[1]}), \texttt{CSR[2]}, and \texttt{wr\_en} (\texttt{CSR[3]}, \texttt{uv\_rst}) are asserted with a one-clock-cycle delay between each signal. The above-mentioned control signals become high for $\frac{n}{2} \times logn$. As shown in Fig.~\ref{fig:shift_reg}, to monitor the control flow integrity of our NTT, a Right Shift Register ($RSR$) is designed similarly to $CRS$. 
Unlike the duplicate $CSR$ in \cite{jati}, our $RSR$ is independent of the main NTT logic. Thus, any tampering in the main logic affects the duplicate $CSR$, but not our $RSR$.
The $3^{rd}$ (msb) bit of this $RSR$ becomes '0' if \texttt{rd\_en} is '0', and it becomes '1' if \texttt{rd\_en} is '1'. Here, the value of this $RSR$ should match the value of the $CSR$ register for the  $\frac{n}{2} \times logn$ clock cycles.
 
\subsubsection{$CFI$ Fault Detector}
The $CFI$ fault detecter consist the $RSR$ and  it compute four fault flags
(1) \texttt{barrett\_cfi} \texttt{\_fault} for $Barrtett$.
(2) \texttt{polymem\_cfi} \texttt{\_fault} for $poly\_mem$.
(3) \& (4) \texttt{uv\_cfi} \texttt{\_fault} for  $uv\_add$ and $uv\_sub$. The final $cfi\_fault$ is computed by performing an AND operation on the aforementioned four fault signals.\\
\textbf{$barrett\_cfi\_fault$: }
The control flow faults in the $barrett$ block can be identified by analyzing the interrelationships between its input control signals (\texttt{barrett\_strt}, \texttt{barrett\_rst}) and output status signal (\texttt{barrett\_done}). In our $NTT$, the behavior of the $barrett$ is considered normal if the following four strict conditions are satisfied:
(i) The \texttt{barrett\_strt} control signal is inversely aligned with the \texttt{barrett\_rst} signal, i.e., $barrett\_strt = not~barrett\_rst$.
(ii) The $barrett$ block activates one cycle after \texttt{rd\_en} goes high (flagged by \texttt{CSR[2]}) and deactivates one cycle before \texttt{wr\_en} goes low (flagged by \texttt{CSR[1]}). Therefore, \texttt{barrett\_strt}=\texttt{CSR[1]} OR \texttt{CSR[2]}.
(iii) For the same arguments mentioned in (iii) \texttt{barrett\_strt}=\texttt{RSR[1]} OR \texttt{RSR[2]}.
(iv) The $barrett$ used in our {\sf Secure NTT} has $2$ pipeline stages. The \texttt{barrett\_done} transitions two cycles after \texttt{barrett\_strt}, aligning precisely with \texttt{CSR[0]}/ \texttt{wr\_en}. This alignment must hold during NTT operation.
\[
\hspace{-0.5em}
\small
\begin{array}{l}
    \texttt{barrett\_cfi\_fault} =  \texttt{'0'}~\text{when} 
     \texttt{~barrett\_strt} =\\
     \quad \texttt{NOT}~\texttt{barrett\_rst} \texttt{AND}~(\texttt{CSR[1]}~\texttt{OR}~\texttt{CSR[2]} = \texttt{'1'})  \texttt{~AND} \\
    \quad (\texttt{RSR[1]}~\texttt{OR}~\texttt{RSR[2]} = \texttt{'1'}) 
     \texttt{AND}~\texttt{barrett\_done} = \texttt{wr\_en}
\end{array}
\]

\textbf{$polymem\_cfi\_fault$: }
The control flow integrity of $poly\_mem$ is considered normal if the control input signals \texttt{wr\_en}, \texttt{rd\_en} and \texttt{poly\_mem\_ce} are strictly aligned with corresponding bits of the $RSR$. The aforementioned control signals of $poly\_mem$ are generated from the $CSR$. (i)\texttt{wr\_en}=\texttt{CSR[0]}, (ii)\texttt{rd\_en}=\texttt{CSR[3]} and (iii)\texttt{poly\_mem\_ce}=\texttt{CSR[0]~OR~CSR[3]}. 
\[
\hspace{-0.5em}
\small
\begin{array}{l}
\texttt{polymem\_cfi\_fault} = '0' \texttt{ when } (\texttt{rd\_en} = \texttt{RSR[3]}) \land {} \\
\quad (\texttt{wr\_en} = \texttt{RSR[0]}) \land (\texttt{poly\_mem\_ce} = \texttt{RSR[0]} \lor \texttt{RSR[3]})
\end{array}
\]

\begin{figure}[!htbp]
\centering
\vspace{-5pt}
\includegraphics[width=0.5\textwidth]{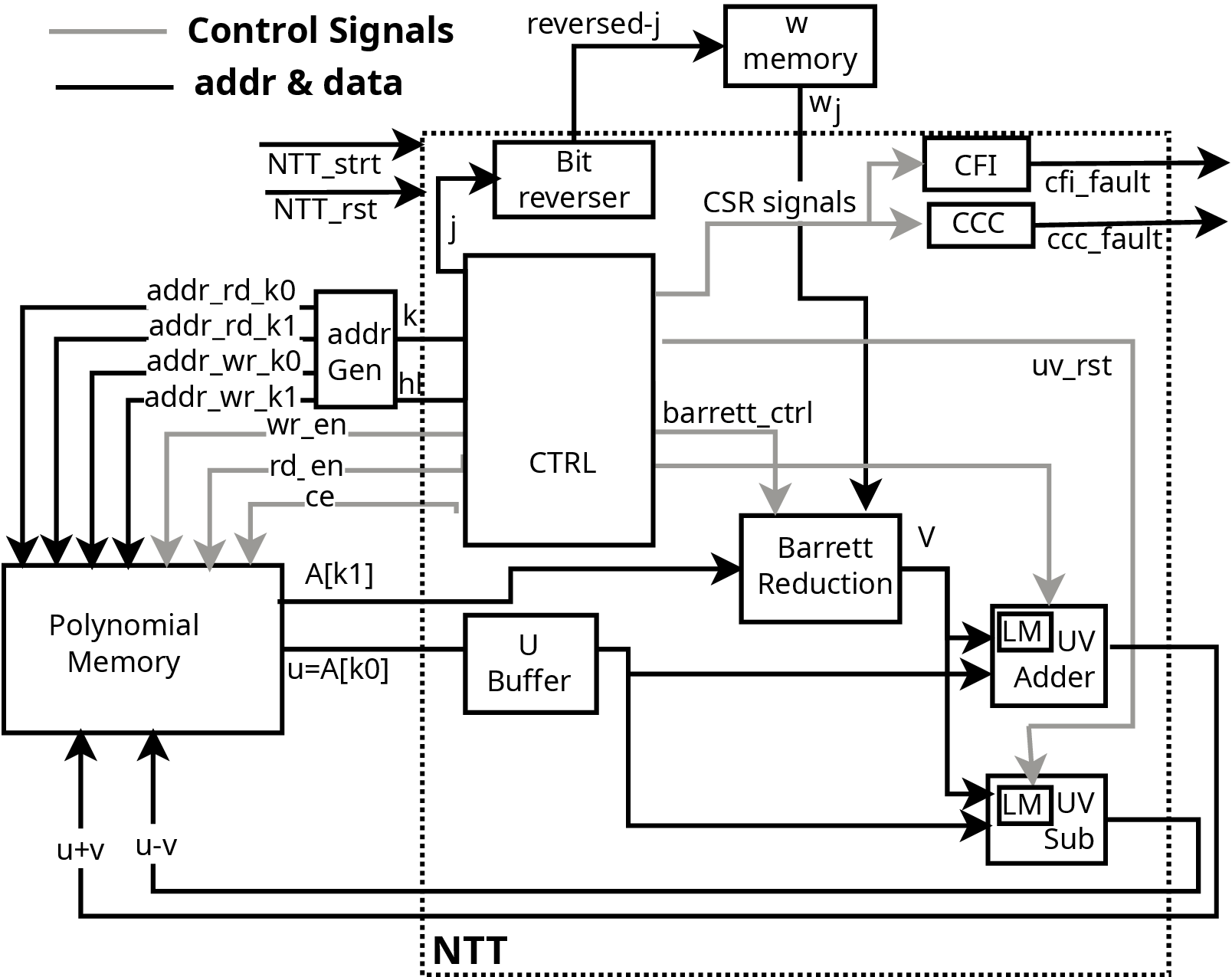}
\vspace{-5pt}
\caption{Architecture of Secure NTT with Fault Detection Module}
\vspace{-10pt}
\label{fig:NTT_arch}
\end{figure}

\textbf{$uv\_cfi\_fault$: }
The $uv\_add$ and $uv\_sub$ blocks are executed in parallel. The alignment of the \texttt{uv\_rst} and \texttt{uv\_strt} input control signals for both $uv\_add$ and $uv\_sub$ is completely identical. Therefore, the proposed {\sf Secure NTT} feeds a common \texttt{uv\_rst} and \texttt{uv\_strt} control signal to both blocks. Here the \texttt{uv\_rst}= NOT~\texttt{CSR[3]} and \texttt{uv\_strt}= \texttt{CSR[0]}. 
\[
\vspace{-2pt}
\hspace{-1.5em}
\small
\begin{array}{l}
    \texttt{uv\_cfi\_fault = '0'~when~(CSR[3]=RSR[3])} \land\\  
                     \quad  \quad \quad \quad  \texttt{(CSR[0] = RSR[0])} \land \texttt{(uv\_strt =} \lnot\texttt{uv\_rst)}
\end{array}
\]
As shown in Fig. \ref{fig:NTT_arch}, the three $CFI$ fault signals belonging to the CSR signals — \texttt{barrett\_cfi\_fault}, \texttt{polymem\_cfi\_fault}, and \texttt{uv\_cfi\_fault} — are connected to the $CFI$ block to generate the final output signal, \texttt{cfi\_fault}.

 \subsection{Clock Cycle Counter ($CCC$)}
Our five-stage pipelined {\sf Secure NTT} requires $(\log n \times \frac{n}{2}) + 4$ clock cycles. For $n = 1024$, our NTT requires 1028 clock cycles for the entire polynomial transformation. The active-high signals - \texttt{rd\_en}, \texttt{wr\_en}, \texttt{polymem\_ce}, \texttt{barrett\_strt}, \texttt{barrett\_done}, and \texttt{uv\_strt} - must remain high or low for a correlated  value of $1024$ clock cycles, starting after \texttt{CTRL\_rst} goes low. Our {\sf Secure NTT} uses a Clock Cycle Counter ($CCC$) to ensure that the $Poly\_mem$, $Barrett$, $uv\_add$, and $uv\_sub$ modules operate for the correct number of clock cycles. Any unintentional or injected faults via hardware Trojans that cause additional clock-cycle delays in any of the aforementioned sub-components can be detected by the $CCC\_fault$ output (shown in Fig. \ref{fig:NTT_arch}) from the $CCC$ module.

\subsection{Local Mask (LM) Unit}
To mitigate SASCA \cite{primas} on our {\sf Secure NTT}, we employ local masking by randomizing the twiddle factors $\omega$ during the write operations to $poly\_mem$. The availability of $\omega$ in $w\_mem$ makes this process convenient.
As shown in Eq. \ref{equ:uv}, the $uv\_adder$ and $uv\_sub$ of the {\sf Secure NTT}  writes $\boldsymbol{\overline{A}}[k0]$ and $\boldsymbol{\overline{A}}[k1]$ in the $poly\_mem$ (Line \ref{uplusv} \& Line \ref{uminusv} of Algorithm \ref{algo:intt}). 
\begin{equation}
\label{equ:uv}
\boldsymbol{\overline{A}}[k0] \gets (U + V) \% q,
~\boldsymbol{\overline{A}}[k1] \gets (U - V) \% q \\
\end{equation}
\vspace{-10pt}
\begin{equation}
\label{equ:uv2}
\boldsymbol{\overline{A}}[k0] \gets (U + V)\omega_r \% q,
~\boldsymbol{\overline{A}}[k1] \gets (U - V)\omega_r \% q \\
\end{equation}
Instead of Eq. \ref{equ:uv}, we compute Eq. \ref{equ:uv2} within the $LM$ units of $uv\_adder$ and $uv\_sub$ during the $poly\_mem$ write operation. Here $\omega_r$ is a random twiddle factor available in $w\_mem$. In inverse NTT, we again multiply $\omega_r^{-1}$ available in $w\_mem$ to get the original polynomial coefficients.

\section{Adaptive Fault Correction Module}
\label{sec:fc}
Our adaptive fault correction methods in NTT, generates $m$ PR bitstream files for the same NTT architecture placed at different locations on the FPGA floor. For example, $PR_1$ corresponds to $NTT_1$, $PR_2$ corresponds to $NTT_2$, and so on, up to $PR_m$ for $NTT_m$.
\subsection{Measures}
If $cfi\_fault$ or $ccc\_fault$ occurs, the {\sf Secure NTT} can perform three types of measures in our architecture:
\subsubsection{Repeat Previous Loop} 
\label{sec:m1}
If a fault is detected in an FPGA, migrating the hardware task to a different FPGA floor or reconfiguring it with a new bitstream is a widely adopted technique. However, hardware task migration, often termed a context switch of the hardware task, is very expensive in terms of timing overhead \cite{context}. Therefore, splitting the NTT flow at specific boundaries or nodes can reduce the context-switching overhead of hardware tasks. Our {\sf Secure NTT} supports fine-grained splitting at each clock cycle. 
If the fault is detected, the $results$ ($U+V$ and $U-V$) of the previous NTT iteration can be discarded and the loop can be re-executed using the same $\omega_j$ from $w\_mem$ and $A[k0]$ and $A[k1]$ from $poly\_mem$ (addressed by $j$ and $k$ respectively) with the same PR bitstream file of the NTT.
\subsubsection{Reload PR Bit \& Repeat Previous Loop}
\label{sec:m2}
The same PR bitstream of the NTT can be reoloaded in the same location of the FPGA floor with the help of Internal Configuration Access Port (ICAP) and repeat measure \ref{sec:m1}.
\subsubsection{Relocate PR Bit \& Repeat Previous Loop}
\label{sec:m3}
The location of the NTT on the FPGA floor can be relocated by using different PR bitstream files of the NTT with the ICAP and repeat measure \ref{sec:m1}.
\par It is to be noted that hardware task context switching requires both reading and writing of bitstreams, whereas splitting the NTT flow eliminates the bitstream reading step, making \ref{sec:m2} and \ref{sec:m3} more lightweight compared to conventional context switching.
The above-mentioned measures are taken based on two thresholds for the $cfi\_fault$ and the $ccc\_fault$ occurrence. For $cfi\_fault$, the threshold values are defined as $cfi\_th\_reld$ and $cfi\_th\_relc$. As shown in line \ref{line:m2cfi} of Algorithm \ref{algo:ntt:correction}, if the $cfi\_fault$ count ($ncfi\_fault$) exceeds $cfi\_th\_reld$ but remains below $cfi\_th\_relc$, then measure \ref{sec:m2} is applied. As shown in line \ref{line:m3cfi} of Algorithm \ref{algo:ntt:correction}, if $ncfi\_fault$ exceeds $cfi\_th\_relc$, then measure \ref{sec:m3} is applied. This check on $ncfi\_fault$ is performed in every NTT loop iteration during each clock cycle.
After completing an NTT iteration, the same procedure is applied to $ccc\_fault$. These steps are detailed in lines 20–27 of Algorithm \ref{algo:ntt:correction}.
To incorporate measures \ref{sec:m1}, \ref{sec:m2} and \ref{sec:m3}, our fault correction module introduces three additional components into the NTT architecture.
\begin{algorithm}
  \caption{Fault Correction of NTT }
  \label{algo:ntt:correction}
  \begin{algorithmic}[1]
    \State \textbf{Input:} $A(x)$, $\omega$, $q$
    \State \textbf{Output:} $\overline{A}(x)$
    \State $hl=\frac{n}{2}$
    \For{$i = 0$ to $log_2(n)-1$}
      \For{$j = 0$ to $2^i - 1$}
        \For{$k = 0$ to $hl-1$ }
         \If{cfi\_fault}
            \State $ncfi\_fault++$
            \If{\parbox[t]{.6\linewidth}{($ncfi\_fault > cfi\_th\_reld$) \& \\ ($ncfi\_fault < cfi\_th\_relc$)}} \label{line:m2cfi}
                 \State  Reload PR bit 
                 \State  Repeat Previous Loop
            \ElsIf{$ncfi\_fault > cfi\_th\_relc$}\label{line:m3cfi}
                  \State Relocate PR bit
                  \State  Repeat Previous Loop
            \Else \label{line:m1cfi}
                \State Repeat Previous Loop
            \EndIf
         \EndIf
        \EndFor
      \EndFor
      \State $hl=\frac{hl}{2}$
    \EndFor
         \If{ccc\_fault}
            \State $nccc\_fault++$
           \If{\parbox[t]{.6\linewidth}{($nccc\_fault > ccc\_th\_reld$) \&  ($nccc\_fault < ccc\_th\_relc$)}}
                 \State  Reload PR bit 
                 \State  Repeat Previous Loop
            \ElsIf{$nccc\_fault >  ccc\_th\_relc$}
                  \State Relocate PR bit
                  \State  Repeat Previous Loop
            \Else
                \State Repeat Previous Loop
            \EndIf
         \EndIf
    \State \Return $\overline{A}$
  \end{algorithmic}
\end{algorithm}
\subsection{Bit Patcher}
The bit patcher is an application that runs on the host CPU.
The primary responsibility of our bit patcher is to manage the PR bitstreams of the NTT. Depending on the required measures, it may reconfigure the same PR bitstream or, in some cases, load a different PR bitstream of the NTT. When a different bitstream is required, our bit patcher selects it based on a risk factor ($R_i$).
\subsubsection{$R_i$ Calculation}
The value of $R_i$ is calculated based on the number and types of faults that are considered for the system.
Suppose a total of $\phi$ fault types are considered, denoted by ${F_1, F_2, \ldots, F_{\phi}}$, with corresponding weights ${W_{F_1}, W_{F_2}, \ldots, W_{F_{\phi}}}$. For any $i^{th}$ PR bit stream $PR_i$ of a component or core, let the number of $F_1$-type faults be $nF_{1_j}$, the number of $F_2$-type faults be $nF_{2_j}$, and so on. If the total number of runs for the $i^{th}$ component or core is $NR_i$, then the expression for calculating $R_i$ can be written as:
\begin{equation}
  R_i = \sum_{j=1}^{\phi} W_{F_j} \cdot \frac{\mathrm{nF}_j/ NR_j}{\mathrm{max\_nF_j}_{k}/NR_{k}} 
  \label{equ:ri}
\end{equation}
Here, $\mathrm{max\_nF}_{jk}$ denotes the maximum number of $F_j$-type fault occurrences for the PR bitstream $PR_k$ in the FPGA floor $F_k$, and $NR_k$ represents  the number of runs of the PR bitstream $PR_k$. The value of $k$ will be different for each loop of $j$. 
 It is possible that a particular PR bistream of a component or core is selected by the bit patcher more frequently than others. Consequently, a  component or core that is invoked more often is also likely to exhibit a higher absolute number of fault occurrences, not necessarily because it is more fault-prone but simply due to its higher utilization. To mitigate this bias, we normalize the raw fault counts by the number of times each  component or core is selected. Specifically, instead of directly using $nF_j$, we consider the normalized quantities $max\_nF_{jk}/NR_k$. The same normalization is also required in the denominator of Eq.~\ref{equ:ri}, where the maximum numbers of $nF_j$ faults occur in $k^{th}$ core or component. Without this adjustment, the denominator would similarly be biased by the number of times a particular core or component is selected. 

The  $\sum_{j=1}^{\phi} W_{F_j}=1$. If $R_i$ is the same for two PR bitstreams, the bit patcher selects 
the PR bitstream with the higher number of runs.
\subsubsection{$R_i$ Calculation for Secure NTT}
In our \textsf{Secure NTT}, we considered only two types of faults:
$cfi\_fault$ and $ccc\_fault$; therefore, $\phi = 2$. Therefore, our bit patcher
collects the number of $cfi\_fault$ and $ccc\_fault$ occurrences from all $m$ available NTTs in our architecture. Note that only one NTT is active at a time, while the remaining $m-1$ NTTs are blank, meaning their PR bitstreams are not configured. The bit patcher application profiles the $m$ NTTs based on the number of $cfi\_fault$ and $ccc\_fault$ occurrences, and computes a risk factor $R_i$ for each NTT. Based on the calculated $R_i$ values, the bit patcher selects the appropriate PR bitstream during measure \ref{sec:m3} only. We have $m$ PR bitstreams $\{PR_1,PR_2,\ldots,PR_m\}$ for $\{NTT_1,NTT_2,\ldots,NTT_m\}$  mapped to FPGA floors
$\{f_1,f_2,\ldots,f_m\}$. 
Each $PR_i$ exhibits runtime fault counts: $\mathrm{ncfi\_fault}_i$, representing the number of $cfi\_fault$ occurrences, and $\mathrm{nccc\_fault}_i$, representing the number of $ccc\_fault$ occurrences for $NTT_i$, where $i \in {1,2,\ldots,m}$. 
In the $R_i$ calculation for $NTT_i$, $\mathrm{ncfi\_fault_i}$ and $\mathrm{nccc\_fault_i}$ are normalized by dividing them by $\mathrm{max\_ncfi\_fault_{j}}$ and $\mathrm{max\_nccc\_fault_{k}}$, respectively. Here, it is assumed that in $NTT_j$ and $NTT_k$, our proposed fault detection module detects the maximum number of $cfi\_fault$ and $ccc\_fault$, respectively. The $\mathrm{max\_ncfi\_fault_{j}}$ denotes the maximum number of $cfi\_fault$s detected in $NTT_j$ and $\mathrm{max\_nccc\_fault_{k}}$ represents maximum number of $ccc\_fault$s detected in $NTT_k$. 
The objective of the bit patcher is to select the PR file with minimal risk where risk $R_i$ depends on both $\mathrm{ncfi\_fault}_i$ and $\mathrm{nccc\_fault}_i$.
We redefine a composite risk factor $R_i$ for $PR_i$ bitstream as
\begin{equation}
  R_i = W_{cfi} \cdot \frac{\frac{\mathrm{ncfi\_fault}_i}{ NR_i}}{\frac{\mathrm{\max\_ncfi\_fault}_{j}}{NR_j}} + W_{ccc} \cdot \frac{\frac{\mathrm{nccc\_fault}_i}{NR_i}}{\frac{\mathrm{max\_nccc\_fault_{k}}}{NR_k}}
  \label{equ:rintt}
\end{equation}
\begin{figure}[!htbp]
\centering
\vspace{-5pt}
\includegraphics[width=0.35\textwidth]{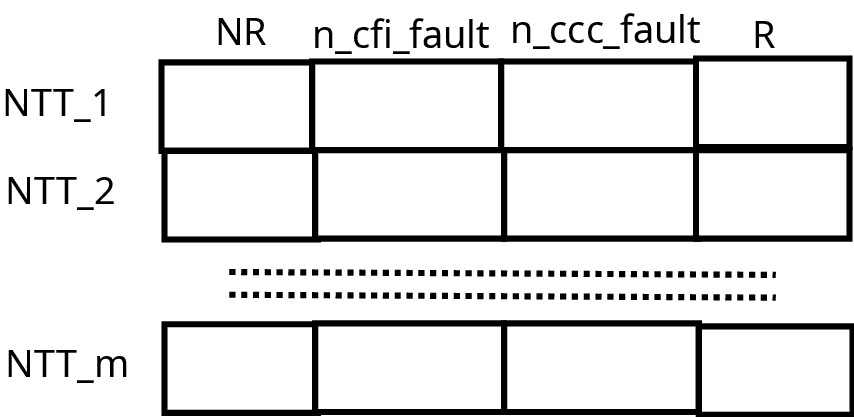}
\vspace{-5pt}
\caption{Bit Patcher Table}
\vspace{-10pt}
\label{fig:bit_patcher}
\end{figure}
Here $W_{cfi}$ and $W_{ccc}$ are weights which reflects the relative importance of control-flow faults $cfi\_fault_i$ and unconventional delays fault $ccc\_fault_i$ of $i^{th}$ NTT which is known as $NTT_i$.
If control-flow faults $cfi\_fault$ are more critical, set $W_{cfi} > W_{ccc}$.
If unconventional delays fault $ccc\_fault$ are more critical, set $W_{ccc} > W_{cfi}$. In our system, we use $W_{cfi} = 0.5$ and $W_{ccc} = 0.5$. 
For the relocation of the NTT in measure~\ref{sec:m3}, 
the bit patcher selects the $PR_i$ bitstream of $NTT_i$ 
that has the minimum risk score $R_i$. As shown in Fig. \ref{fig:bit_patcher}, the bit patcher maintain a table which store number of runs ($NR$),  $ncfi\_fault$, $nccc\_fault$ and $R$ for each NTT.
\begin{figure}[!htbp]
\centering
\vspace{-5pt}
\includegraphics[width=0.5\textwidth]{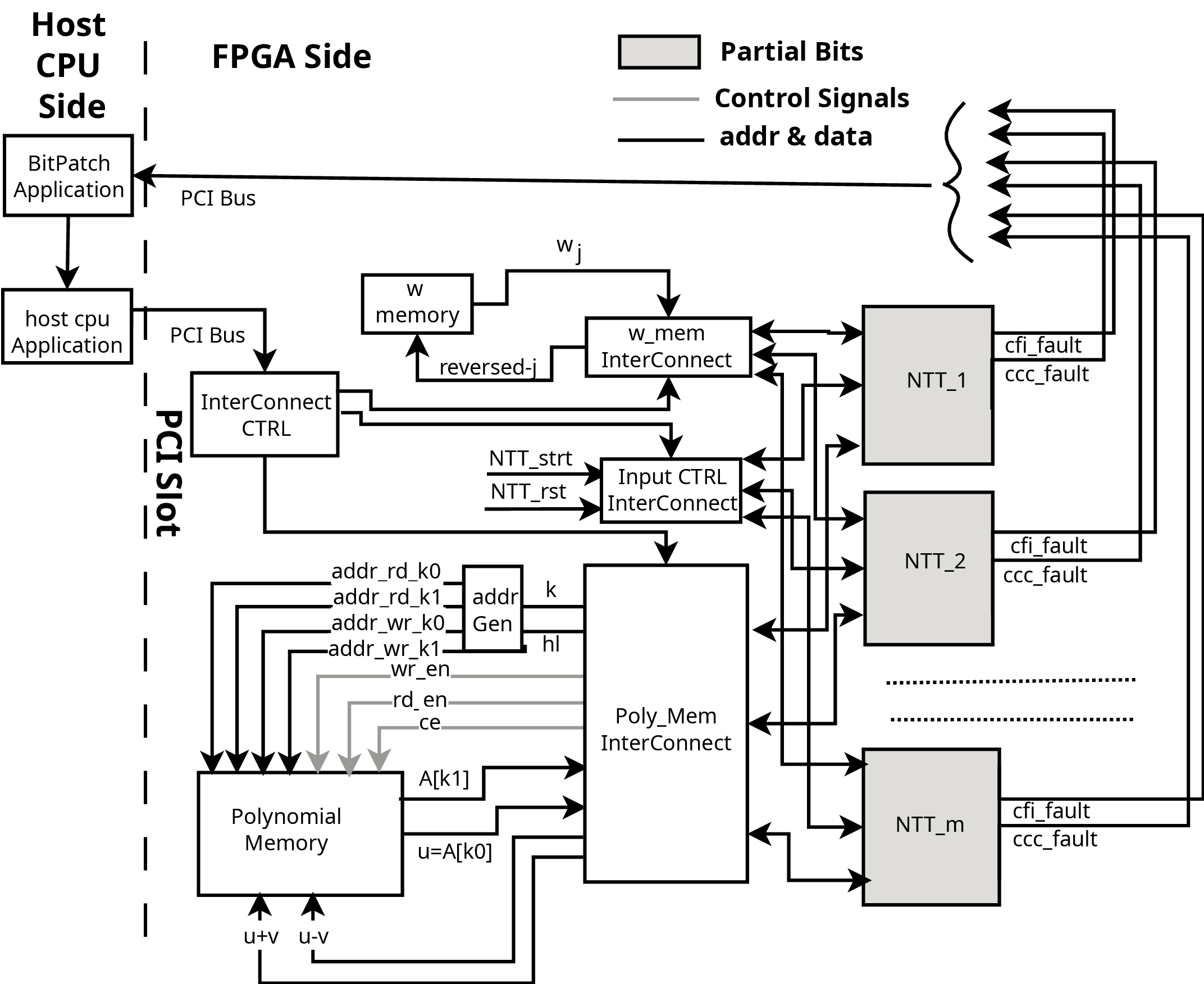}
\vspace{-5pt}
\caption{Architecture of Secure NTT With Fault Correction Module}
\vspace{-10pt}
\label{fig:NTT_arch:cd}
\end{figure}

\subsection{Bus Inter Connects}
\label{sec:bus}
Except for the clock input, our NTT module has four inputs: $NTT\_rst$, $NTT\_strt$, $A[k0]$, and $A[k1]$, and eight outputs: $j$, $k$, $hl$, $rd|en$, $wr\_en$, $ce$, $U+V$, and $U-V$. The NTT only interacts with $poly\_mem$ and $w\_mem$. Our fault correction methods allow a fresh NTT to be loaded in a different FPGA floor. For this reason, we define multiple NTT instances across FPGA floors. Whenever a new NTT is loaded, it must be connected to $poly\_mem$ and $w\_mem$. Therefore, the connections between $poly\_mem$, $w\_mem$, and the NTT must be dynamic. To achieve this, we create three bus interconnects that link control inputs ($NTT\_rst$, $NTT\_strt$), $poly\_mem$ and $w\_mem$ to the currently active NTT. At any given time, only one NTT is active, while the remaining NTTs remain unconfigured (their PR bitstreams are not loaded).
In our fault detection method, we use three bus interconnects: 
(i) \textit{$poly\_mem$ InterConnects}, which connects $poly\_mem$ to the selected NTT, 
(ii) \textit{$w\_mem$ InterConnects}, which connects $w\_mem$ to the selected NTT, and 
(iii) \textit{Input CTRL InterConnects}, which connects the control inputs 
($NTT\_rst$, $NTT\_strt$) to the selected NTT.
Whenever bitpatcher relocate NTT in different FPGA floor, bit patcher configure the bus interconnects using the $InterConnect~CTRL$.For all three bus interconnects, we use a buffer to store the previous values. This is necessary because if a fault is detected, the NTT must recompute the previous loops using these buffered values. The proposed {\sf Secure NTT} along withe fault correction module is shown in Fig. \ref{fig:NTT_arch:cd}. The $addr\_gen$ takes $k$ and $hl$ from NTT and generate read and write addresses of $poly\_mem$.
\begin{figure}[!htbp]
\centering
\vspace{-5pt}
\includegraphics[width=0.5\textwidth]{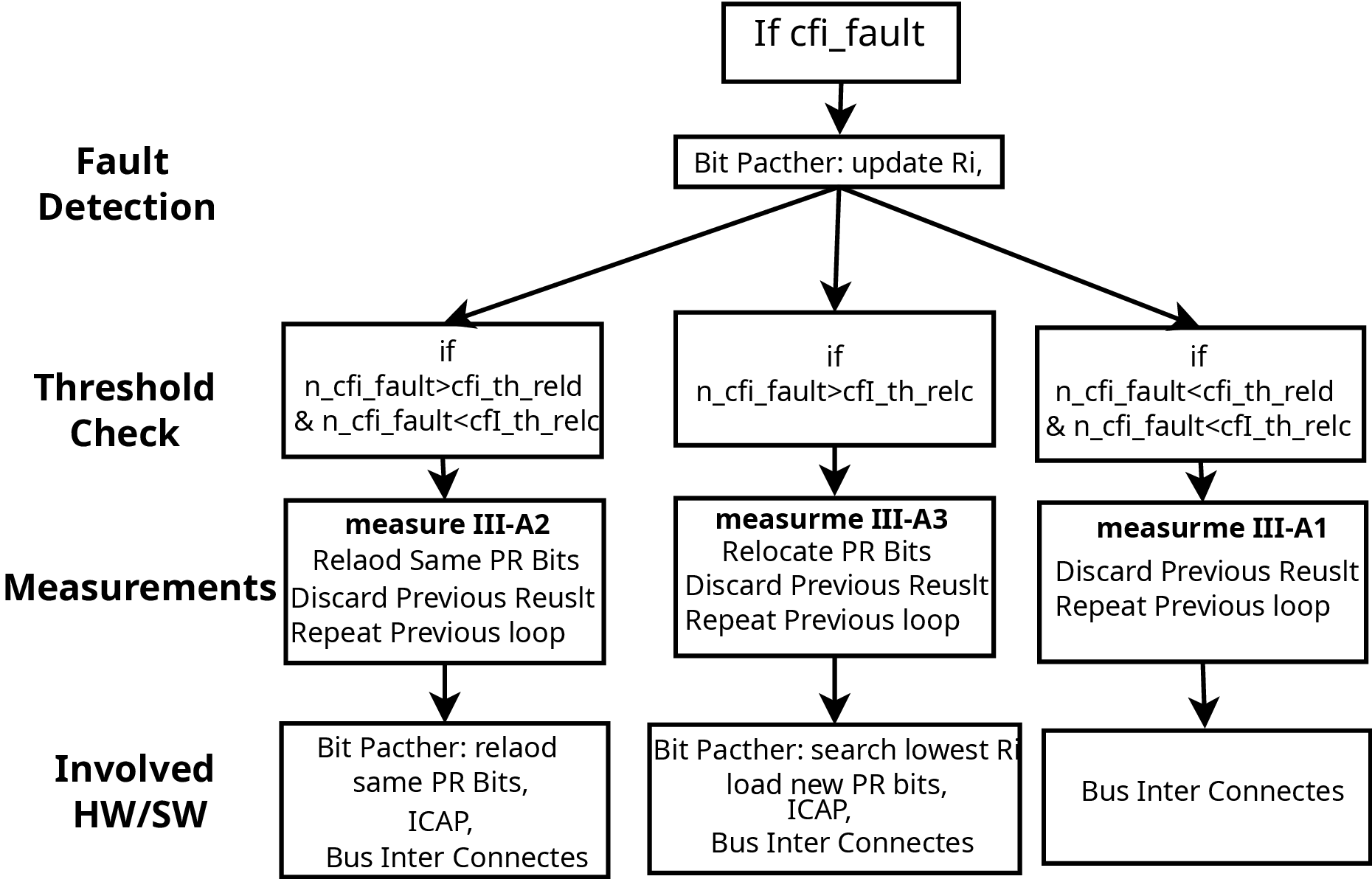}
\vspace{-5pt}
\caption{Flow Diagram of Fault Correction Module for $cfi\_fault$}
\vspace{-10pt}
\label{fig:fc_flow}
\end{figure}
\subsection{Fault Correction Flow}
Whenever the proposed $cfi\_fault$ and $ccc\_fault$ mechanisms detect faults, they update the fault counts to the bit patcher, which consequently updates the $R_i$ of the corresponding NTT block. Thereafter, three conditions stated in Algorithm \ref{algo:ntt:correction} are checked based on four thresholds: $cfi\_th\_reld$, $cfi\_th\_relc$, $ccc\_th\_reld$, and $ccc\_th\_relc$. The three conditions for $cfi\_fault$ occurrences, shown in line \ref{line:m2cfi} as measure \ref{sec:m2}, line \ref{line:m3cfi} as measure \ref{sec:m3}, and line \ref{line:m1cfi} as measure \ref{sec:m1} of Algorithm \ref{algo:ntt:correction}, trigger the three types of fault corrections implemented in our {\sf Secure NTT}. The proposed fault correction module involves only the bus interconnects (stated in Sec. \ref{sec:bus}) for measure \ref{sec:m1}. It involves reloading same PR bit step of bit pacther, ICAP \cite{xilinx_icap} and bus interconnects for measure \ref{sec:m2}. Finally for measure \ref{sec:m3}, it involves searching for lowest $R_i$ NTT, download new PR bit steps of Bit patcher, ICAP and bus interconnects. Fig.~\ref{fig:fc_flow} shows the four steps that are performed following the detection of a $cfi\_fault$. Depending on the three types of measures, the hardware and software components involved in each step differ accordingly. The same sequence of steps is also applied when a $ccc\_fault$ is detected.

\section{Implementation \& Results}
\label{sec:rai}
As depicted in Fig. \ref{fig:NTT_arch:cd}, the proposed {\sf Secure NTT} architecture is implemented on an Artix-7 FPGA using VHDL and the Vivado 2022.2 tool. The host side code is written in C. The NTT is placed inside FPGA is connected with a $i5$ host CPU through PCI bus. 
\subsection{Our Validation Strategies}
To validate our fault detection and correction methods, we implemented three variants of Kyber (Kyber-512, Kyber-768 and Kyber-1024 ) with four NTT PR bitstreams. In Kyber variants, the parameter $n$ (number of coefficients) is $256$, and the modulus $q$ is $3329$. Each coefficient requires $12$ bits, so the size of the \texttt{poly\_mem} memory is $12 \times 256$. The \texttt{w\_mem} memory, which stores the twiddle factors, is also $12 \times 256$. Both the NTT and the entire Kyber implementation run at a $100$~MHz clock frequency. We assume a system setup consisting of an Intel i5 host CPU and an Artix-7 FPGA, which together act as a server platform. In this scenario, the server is expected to execute the Kyber Key Generation, Encapsulation, and Decapsulation processes multiple times, as would be required in practical secure communication applications.

\begin{table*}[ht]
\centering
\renewcommand{\arraystretch}{1.2} 

\begin{tabular}{|p{1.0cm}|p{0.8cm}|
                p{0.8cm}|p{1.5cm}|p{1.2cm}|
                 p{0.8cm}|p{1.5cm}|p{1.2cm}|
                p{0.8cm}|p{1.5cm}|p{1.2cm}|
                }
\hline
\multirow{2}{*}{\parbox{1.0cm}{\centering\textbf{Block}}} & 
\multirow{2}{*}{\parbox{0.8cm}{\centering\textbf{Sample Size}}} & 
\multicolumn{3}{c|}{\textbf{Kyber-512}} & 
\multicolumn{3}{c|}{\textbf{Kyber-768}} & 
\multicolumn{3}{c|}{\textbf{Kyber-1024}} \\ \cline{3-11}

& & \textbf{\#NTT Run / Block} & \textbf{Total \# NTT Runs \& \# Faults Injected} & \textbf{Detection \& Correction Eff. (\%)} & 
      \textbf{\#NTT Run / Block} & \textbf{Total \# NTT Runs \& \# Faults Injected} & \textbf{Detection \& Correction Eff. (\%)} & 
      \textbf{\#NTT Run / Block} & \textbf{Total \# NTT Runs \& \# Faults Injected} & \textbf{Detection \& Correction Eff. (\%)} \\ \hline

\parbox{1.0cm}{\centering\textbf{KeyGen}} & \parbox{0.8cm}{\centering 262144} & 
  2 & 524288 & 100 & 3 & 786432 & 100 &  4 & 1048576 &100 \\ 
&&2 & 524288 & 100& 3 & 786432 & 100 &  4 & 1048576 &100  \\ 
\hline

\parbox{1.0cm}{\centering\textbf{EnCap}} & \parbox{0.8cm}{\centering 262144} & 
  2 & 524288 & 100 &  3 & 786432 &100  & 4 & 1048576 & 100 \\
&&2 & 524288 &100  & 3 & 786432 & 100 &  4 & 1048576 & 100 \\ 
&&1 & 262144 & 100 & 1 & 262144 &100  &  1 & 262144 &100  \\ \hline

\multirow{2}{*}{\parbox{1.0cm}{\centering\textbf{DeCap}}} & 
\multirow{2}{*}{\parbox{0.8cm}{\centering 262144}} & 
2 & 524288 & 100 &  3 & 786432 & 100 &  4 & 1048576 & 100 \\
&&2 & 524288 & 100 & 3 & 786432 &  100&  4 & 1048576 &  100\\ 
&&2 & 524288 & 100 & 3 & 786432 &  100&  4 & 1048576 & 100 \\ 
&&1 & 262144 & 100 & 1 & 262144 & 100 &  1 & 262144 & 100 \\ 
&&1 & 262144 & 100 & 1 & 262144 &  100&  1 & 262144 &  100\\ 
\hline

\textbf{Total} & \textbf{12,288} &  17 & \textbf{44,56,448}  &  \textbf{100} & 24  &  \textbf{62,91,456} &  \textbf{100} & 31 & \textbf{81,26,464}& \textbf{100} \\ \hline

\end{tabular}
\caption{Fault Detection and Correction Efficiency for Emulated Faults by Our Fault Injector for Kyber Variants}
\label{tab:fault_kyber}
\end{table*}

\begin{figure*}[!htbp]
\centering
\vspace{-5pt}
\includegraphics[width=0.8\textwidth]{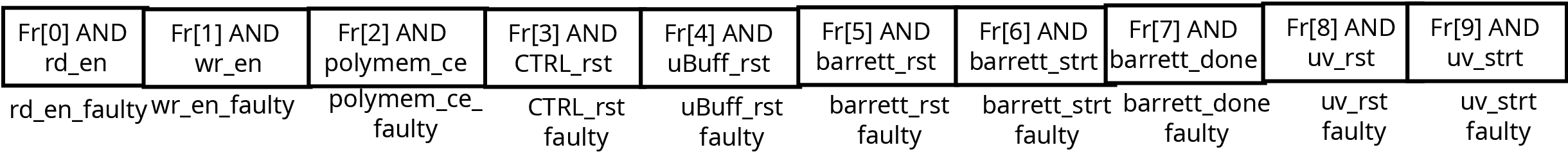}
\vspace{-5pt}
\caption{Connection of Fault Injector with Control Signals}
\vspace{-10pt}
\label{fig:fr}
\end{figure*}

During Kyber's \textit{Key Generation}, \textit{Encapsulation}, and \textit{Decapsulation} processes, each NTT operation requires $1024$ clock cycles. 
To emulate the faults caused by a hardware Trojan, we designed a fault injector. This block generates a 10-bit output named as $F_r$, which is connected to the ten $AND$ signals of our {\sf Secure NTT} listed in the $2^{nd}$ column of Table~\ref{tab:csr}. The 10-bit $F_r$ from the fault injector determines whether the actual values of the corresponding signals in Table~\ref{tab:csr} are passed to the NTT or blocked. As shown in Fig. \ref{fig:fr}, instead of using $rd\_en$ directly to read the polynomial memory, the system uses $F_r[0]~AND~ rd\_en$. A similar modification is applied to the other nine control signals.
We generated two random numbers using the special device file \texttt{/dev/urandom} available in our Ubuntu host CPU. These two random numbers, named as $R_t$ and $R_s$ are sent to the fault injector through PCI bus. The first random number $R_t$, selected in the range $[0, 1023]$, determines the specific clock cycle at which the attack is performed on our NTT. The second random number $R_s$, also in the range $[0, 1023]$, is used to feed the 10-bit output $F_r$ of our fault injector, which specifies the control signal to be compromised. For example, if $R_t = 1002$ and $R_s = 766$, then after the activation of our {\sf Secure NTT}, at the $1002^{\text{nd}}$ clock cycle, the value of $R_s$ will be loaded at the injector output $F_r$. In this case, $F_r = \texttt{1011111110}$, which means that the actual values of $rd\_en$ and $uv\_rst$ will not be blocked by our fault injector. Therefore, in our fault emulation process, at a random clock cycle of the NTT, we tamper with the control signals of the NTT, which emulates the behavior of a hardware Trojan attack. 
As shown in Table \ref{tab:fault_kyber}, we have run the Kyber Key Generation, Encapsulation, and Decapsulation processes 262144 times. Since each of the three Kyber processes (Key Generation, Encapsulation, and Decapsulation) requires multiple executions of the NTT/INTT, the total number of NTT/INTT operations depends on the specific Kyber variant. For Kyber-512, Kyber-768, and Kyber-1024, we executed our {\sf Secure NTT} implementation 69{,}632, 98{,}304, and 1{,}26{,}976 times, respectively. In each of these runs, for the entire 1024 clock cycles of NTT activation period, we targeted a specific clock cycle (denoted by $R_t$) and performed 'Stuck-at-1' and 'Stuck-at-0' attacks on random control signals selected through $R_s$. As shown in Table \ref{tab:fault_kyber}, 100\% of the faults emulated by our fault injector are detected by $CFI$ and $CCC$, and all detected faults are successfully corrected by our adaptive fault-correction module. For Kyber-512, Kyber-768, and Kyber-1024, we use eight NTT PR bitstreams with eight different slice ranges specified in the Xilinx Design Constraints (XDC) file. The four threshold values are set to $cfi\_th\_reld = 256$, $cfi\_th\_relc = 512$, $ccc\_th\_reld = 256$, and $ccc\_th\_relc = 512$
The source code of our NTT is available on GitHub \footnote{https://github.com/rourabpaul1986/NTT/tree/master/fntt\_pipelined}.
\subsection{Overheads}
\label{sec:overheads}

The proposed adaptive fault correction scheme and fault detection modules incur a 19.7\% slice overhead and a 3\% energy overhead compared to the baseline NTT. The timing overhead of the fault correction module depends on the types of measures. 
For measure \ref{sec:m1}, the overhead is limited to a single clock period of the NTT engine (10 ns in the present configuration), and the $R_i$ update is performed on the host CPU. Because the update runs on the host and the NTT computation continues on dedicated hardware, the host-side $R_i$ update does not affect the executing NTT pipeline. For measure \ref{sec:m2}, the time overhead is dominated by two factors: (i) Time required for $R_i$ update within the bit patcher logic, and (ii) Time to stream the corresponding PR bitstream through the ICAP. For a PR bitstream file of approximately 56 kB NTT, the ICAP transfer requires on the order of $\sim$ 150 $\mu$s with a 100 MHz ICAP clock and 32-bit data width. This value represents the raw transfer time and excludes any additional protocol framing or controller overhead.
 For measure \ref{sec:m3}, the time overhead is dominated by three factors: (i)Time required for $R_i$ update within the bit patcher logic, and (ii) PR bitstream configuration time by ICAP and (iii) Time to choose the PR bit stream based $R_i$, which takes around $\sim$ 256 $\mu$s.

\subsection{Different Hardware Trojan Scenarios}
As shown in Table \ref{tab:trojan-ntt}, for our {\sf Secure NTT}, we consider four scenarios based on possible hardware Trojan locations in the system: Outside Upstream, Inside NTT, Outside Downstream, and Inside Monitors. We also consider four different phases of hardware Trojan insertion in the FPGA: RTL, Synthesis/Foundry, Bitstream level, and Post-deployment. Based on the literature, it can be concluded that at the RTL and Synthesis/Foundry phases, all four Trojan locations are possible. At the Bitstream level, attacks on Outside Upstream/Downstream and fault-detection modules are feasible, while Inside-core attacks are harder but still possible. In the Post-deployment phase, only Outside Upstream/Downstream attacks are realistic. Table \ref{tab:trojan-ntt} illustrates how the location of a hardware Trojan impacts our $CFI$, $CCC$, and $LM$.
\subsubsection{Outside Upstream \cite{outup}}
Outside upstream means the hardware Trojan is placed on the control signals before they reach the NTT or target core. In our case, the Trojan might be inserted into the data signals coming from $poly\_mem$ ($A[k0]$, $A[k1]$) or from $w\_mem$ ($w_j$). These types of Trojans can perform a SASCA-style attack through raw inputs. However, the data signals in the first NTT iteration may arrive without masking, while all subsequent outputs ($U+V$ and $U-V$) stored in $poly\_mem$ are written after masking using our $LM$ circuit. Therefore, during the entire NTT activation period, the probability of SASCA-style leakage is only $1/1024$ in our Kyber implementation, whereas for the unprotected existing NTT it is $1$. Pre-masking taps or removal of masking are also not possible at the outside bitstream level, since the $LM$ is placed inside the NTT.
\subsubsection{Inside NTT/Target Core \cite{inside}}
The Trojan is embedded within the NTT itself, directly manipulating its internal signals, operations such as butterfly units, modular reductions, or memory accesses. In our NTT, if a hardware Trojan is inserted within the NTT itself and possibly introduced through compromised synthesis or foundry stages \cite{mukhrejee}, control-flow hijacking or trigger-based payloads are effectively detected by $CFI$, while Trojan-induced delays are reliably detected by $CCC$.
\subsubsection{Outside–Downstream \cite{monitor}}
The Trojan resides after the NTT output, i.e., outside the NTT block, and manipulates or taps the results before they are consumed by subsequent cryptographic operations. A SASCA type leakage through output tapping is not effective because the two output signals of our NTT, $U+V$ and $U-V$, are always masked by the $LM$. Furthermore, output manipulation in this case does not fall under control signal attacks, and is therefore considered out of scope.
\subsubsection{Inside Monitors \cite{monitor}}
Another possible scenario arises when the hardware Trojan is inserted inside the fault detection module itself. In our case, if the Trojan compromises either the $CCC$ or $CFI$, the system may fail to detect faults accurately, potentially leading to overall failure of our {\sf Secire NTT}.

\begin{table*}[!htb]
\centering
\caption{Trojan Scenarios for NTT with Local Mask (LM) Unit, Control-Flow Integrity ($CFI$) Checker, and Clock Cycle Counter ($CCC$)}
\label{tab:trojan-ntt}
\renewcommand{\arraystretch}{1.2}
\begin{tabular}{|p{1.1cm}|p{2.2cm}|p{2.0cm}|p{2.4cm}|p{0.6cm}|p{0.7cm}|p{0.7cm}|p{4.4cm}|}
\hline
\textbf{Trojan Location} & \textbf{Trojan Type / Payload} & \textbf{Insertion Phase} & \textbf{Impact on NTT} & \textbf{LM} & \textbf{CFI} & \textbf{CCC} & \textbf{Remarks}\\
\hline
& SASCA-style leakage via raw inputs & Post-deployment \cite{postdep} / side-channel \cite{ravi2}, \cite{rafael} & Secret key/ plaintext leakage  & $\checkmark$ & $\times$ & $\times$& Input data of NTT is masked by $\omega_r$ in $LM$\\\cline{2-8}
Outside Upstream \cite{outup} & Pre-mask tap (copies inputs before masking) & RTL / synthesis & Secret leakage (mask bypass) & $\times$ & $\times$ & $\times$ & Not possible as $LM$ is placed inside NTT\\\cline{2-8}
& Clock glitch/ insertion, Reset spoofing, DoS, loss of state & Post-deployment \cite{postdep}  / fault injection & Integrity fault, timing errors & $\times$ & $\sim$ & $\sim$ & Trojan in the $NTT$ main $clk$ or $rst$ introduces ambiguity in control flow and may cause delays, which can be detected by $CSR$ and $CCC$ respectively\\\hline\hline
 & Data leak after mask removal (tap internal nets) & RTL \cite{dai} / synthesis \cite{mukhrejee} & Secret leakage inside datapath & $\times$ & $\times$ & $\times$& If a hardware Trojan inside the NTT removes $LM$, the attack directly disables a critical functionality of the design. Since the $LM$ has no explicit control pins to manipulate, such an attack cannot be classified as a control-signal attack; rather, it constitutes a payload-level functional disruption Trojan.\\\cline{2-8}
Inside NTT \cite{inside} & Control-flow hijack (illegal state/branch) & RTL \cite{dai} / synthesis \cite{mukhrejee} & Integrity violation, wrong outputs & $\times$ & $\checkmark$ & $\sim$ & Independent shift registers ($CSR$) inside $CFI$ mimic NTT’s control flow; deviations indicate a Trojan. If timing delays in NTT subcomponents (read, write, Barrett, etc.) are detected by $CCC$\\\cline{2-8}
 & Trigger-based payload (rare sequence) &RTL \cite{dai} / synthesis \cite{mukhrejee} & Conditional leakage or DoS & $\times$ & $\checkmark$ & $\sim$ & If a conditionally activated Trojan alters control flow or delays NTT subcomponents, both $CSR$ and $CCC$ can detect it for the reasons stated above\\\cline{2-8}
& Timing/ delay Trojan (extra states, gated clock) & RTL \cite{dai} / synthesis \cite{mukhrejee} & DoS or covert timing side channel  & $\times$ & $\sim$ & $\checkmark$ & If the Trojan is covert but adds extra states or a gated clock, causing NTT stalling or delays, it can be detected by $CCC$ and $CFI$.\\\hline\hline
 & Output manipulation exfiltration (covert I/O pin, encoding) & RTL \cite{dai} / synthesis \cite{mukhrejee} & Secret leakage via outputs & $\times$ & $\times$ & $\times$& It's not control-signal attacks, and is therefore considered out of scope. \\\cline{2-8}
Outside Downstream \cite{monitor} & SASCA-style leakage via raw outputs & RTL \cite{dai} / synthesis \cite{mukhrejee} & Secret leakage via outputs & $\sim$ & $\times$ & $\times$& During the entire NTT operation and subsequent Kyber/Dilithium processing, inputs remain masked and are only unmasked after the inverse NTT (INTT). If output exfiltration occurs after INTT, it cannot be protected by $LM$; otherwise, any leakage before INTT can be mitigated by $LM$ \\\hline \hline
 & Trojan in $CFI$ (forces ``pass'') & RTL \cite{dai} / synthesis \cite{mukhrejee} & Integrity violation undetected (I) & $\times$ & \textbf{\tiny{Tampered}} & $\sim$& $CFI$ Compromized\\\cline{2-8}
Inside monitors \cite{monitor} & Trojan in counter (forges/freeze count) & RTL \cite{dai} / synthesis \cite{mukhrejee} & Timing deviations undetected  & $\times$ & $\sim$ & \textbf{\tiny{Tampered}}& $CCC$ Compromized\\
\hline
\end{tabular}
\end{table*}
\subsection{Comparison with Existing Works}
In the literature, many NTT implementations exist. Based on fault detection characteristics, they can be categorized into two types: those focusing on data-signal fault detection \cite{paul}, \cite{sven}, and those focusing on control-signal fault detection \cite{jati}, \cite{ravi2}, \cite{rafael}. Since faults on NTT data signals are beyond the scope of this paper, we restrict our comparison to works that focus specifically on NTT control signals. Jati et al. \cite{jati} implement an NTT on an Artix-7 FPGA using Random Access Memory and a Clock Cycle Counter ($CCC$) to prevent SASCA and unconventional delays. Ravi et al. \cite{ravi2} and Rafael et al. \cite{rafael} both apply local masking to polynomial coefficients to protect the NTT from SASCA. The key difference is that \cite{ravi2} employs randomly accessible polynomial coefficient memory, which provides additional protection against SCAs. In our fault detection methodology, we adopt existing techniques such as local masking and the clock cycle counter, but additionally introduce a lightweight shift-register–based backup CSR that is fully independent of the NTT. This fault detection modules imposes only 8.7\% slice overhead and a 2\% energy overhead compared to the baseline NTT. It does not cause any timing overhead. Our {\sf Secure NTT} also implements adaptive fault correction methods with three different measures : \ref{sec:m1}, \ref{sec:m2} and \ref{sec:m3}. The proposed adaptive fault-correction and fault-detection modules impose a combined resource overhead of 19.7\% in slices and an energy overhead of 3\% relative to the baseline. The timing overhead varies with the specific measurement method; details are provided in Sec.~\ref{sec:overheads}. These overheads are tolerable in practical implementations of NTT, as used in Kyber, Dilithium, and other lattice-based cryptographic algorithms. As shown in Table \ref{tab:pr_fc_overhead}, we compare our adaptive fault-correction approach with existing PR-based fault-correction techniques. While existing fault correction solutions focus mainly on recovering from configuration or memory upsets with large area or timing penalties, our design uniquely targets control signal faults induced by hardware Trojans and side-channel conditions. The results demonstrate that the overhead introduced by our method remains reasonable compared with current systems. To the best of our knowledge, this is the first fault-correction strategy that adapts its behavior to the nature of the fault itself.

\section{Conclusion}
\label{sec:con}
In this paper, we propose a {\sf Secure NTT} architecture that detects faults using two key techniques: (i) a shift-register-based lightweight backup control status register to ensure control-flow integrity, and (ii) a clock-cycle counter applied to all critical signals to detect unconventional delays. In addition, a local masking scheme is employed to prevent side-channel attacks such as SASCA. Together, these measures prevent control-flow disruptions, hardware Trojan induced delays in subcomponents, and side-channel leakages. We also implement an adaptive fault-correction module for our {\sf Secure NTT}, tailored to the specific fault types caused by hardware Trojans. Our results shows implementation cost of the above mentioned measures in our {\sf Secure NTT} is nominal and competitive with the existing NTTs.

\section{Future Scope}
\label{sec:fs}
The proposed fault detection technique is highly logic-specific to the NTT algorithm; however, the adaptive fault correction methods can also be applied to other critical modules of lattice-based PQC algorithms. In the future, we plan to implement a PQC security processor integrating Kyber for public-key encryption, Dilithium for digital signatures, and AES-256 for symmetric-key encryption. In this security processor, different fault detection modules will be adopted for the critical components of Kyber, Dilithium, and AES—such as Noise Sampling, KECCAK, Matrix–Vector Multiplication, S-Box, ShiftRows, MixColumns, etc.—while all components will share the same adaptive fault correction mechanism as proposed here.\\
\textbf{Acknowledgment}
This publication has emanated from research conducted with the financial support of Taighde Éireann - Research Ireland under Grant number 13/RC/2077\_P2 at CONNECT: the Research Ireland Centre for Future Networks.

\bibliographystyle{unsrt}  
\bibliography{IEEEexample}

\end{document}